\documentclass{acmproc-sp}

\usepackage{fullpage}
\usepackage{amsmath,latexsym,amssymb}
\usepackage{array}
\usepackage{theorem}

\usepackage{algorithm}
\usepackage[noend]{algorithmic}
    \renewcommand{\algorithmiccomment}[1]{\quad \textit{// #1}}
    \renewcommand{\algorithmicrequire}{\textbf{Input:}}
    \renewcommand{\algorithmicensure}[1]{\textbf{Output:}}

\title{Polynomial-Time Approximation Scheme for Data~Broadcast}

\author{
  Claire Kenyon
  \and    Nicolas Schabanel
  \and    Neal Young
}
\date{}

\theoremstyle{plain}

\theoremheaderfont{\normalfont\bfseries}
\theorembodyfont{\normalfont\itshape}
\newtheorem{theorem}{Theorem}
\newtheorem{lemma}{Lemma}

\newtheorem{proposition}{Proposition}
\newtheorem{corollary}{Corollary}

\theorembodyfont{\normalfont}
\newtheorem{definition}{Definition}
\newtheorem{notation}{Notation}
\newtheorem{remark}{Remark}
\newtheorem{note}{Note}

\newenvironment{proofOf}[1]{
    \paragraph*{\quad\normalfont\scshape Proof of {#1}}
} { 
    $\Box$
    \smallskip
}

\newenvironment{proofSketch}{
    \paragraph*{\quad\normalfont\scshape Proof sketch}
} { 
    $\Box$
    \smallskip
}

\newcommand{\paragraphtitle}[1]{\noindent {\itshape #1}}

\renewcommand{\th}{{\ensuremath{^{\text{th}}}}}
\newcommand{\st}{{\ensuremath{^{\text{st}}}}}
\newcommand{\nd}{{\ensuremath{^{\text{nd}}}}}

\newcommand{\ie}{\textit{i.e.}}

\newcommand{\WLOG}{\textit{w.l.o.g.}}

\newcommand{\DS}{\displaystyle}
\newcommand{\CD}{\ensuremath\centerdot}

\newcommand{\CC}{\ensuremath{\mathcal C}}
\newcommand{\event}{\ensuremath{\mathcal E}}

\newcommand{\eqdef}{\ensuremath{=_{\text{\normalfont def}}}}

\newcommand{\series}[4]{{\ensuremath{{#1}_{#2}{#4}\ldots{#4}{#1}_{#3}}}}

\newcommand{\expect}{\ensuremath{\operatorname{\mathbb E}}}

\renewcommand{\epsilon}{\varepsilon}
\renewcommand{\leq}{\leqslant}
\renewcommand{\geq}{\geqslant}

\newcommand{\LB}{\ensuremath{\operatorname{LB}}}
\newcommand{\LBNoC}{\ensuremath{
        \operatorname{LB_{\operatorname{no\,cost}}}}}

\newcommand{\ART}{\ensuremath{\operatorname{ERT}}}
\newcommand{\COST}{\ensuremath{\operatorname{COST}}}
\newcommand{\BC}{\ensuremath{\operatorname{BC}}}
\newcommand{\AVG}{\ensuremath{\operatorname{Avg}}}

\newcommand{\OPT}{\ensuremath{\operatorname{OPT}}}

\newcommand{\Gjk}{\ensuremath{G_{j,k}}}
\newcommand{\gjk}{\ensuremath{g_{j,k}}}

\newcommand{\dP}{{\ensuremath{\dot p}}}
\newcommand{\dC}{{\ensuremath{\dot c}}}
\newcommand{\dS}{{\ensuremath{\dot S}}}
\newcommand{\dM}{{\ensuremath{\dot M}}}

\newcommand{\NP}{\ensuremath{N\!P}}

\begin{document}
\openout15=proofs.deferred

\maketitle




\begin{abstract}\small
  The data broadcast problem is to find a schedule for broadcasting a given set
  of messages over multiple channels.  The goal is to minimize the cost of the
  broadcast plus the expected response time to clients who periodically and
  probabilistically tune in to wait for particular messages.
  
  The problem models disseminating data to clients in asymmetric communication
  environments, where there is a much larger capacity from the information
  source to the clients than in the reverse direction.  Examples include
  satellites, cable TV, internet broadcast, and mobile phones.  Such
  environments favor the ``push-based'' model where the server broadcasts
  (pushes) its information on the communication medium and multiple clients
  simultaneously retrieve the specific information of individual interest.
  This sort of environment motivates the study of ``broadcast disks'' in
  Information Systems \cite{A98,BC96}.
  
  In this paper we present the first polynomial-time approximation scheme for
  the data broadcast problem for the case when $W=O(1)$ and each message has
  arbitrary probability, unit length and bounded cost.  The best previous
  polynomial-time approximation algorithm for this case has a performance ratio
  of $9/8$ \cite{BBNS98}.
\end{abstract}


\section{Background and Result}
The input is a set $M=\{\series M1m,\}$ of messages, each with a probability
$p_i$ and cost $c_i$, and a parameter $W$ --- the number of channels.  The
output is (finitely described) infinite {\em broadcast schedule $S$} for the
messages --- specifying for each time $t=0,1,2,\ldots$ and channel $k$, a
message $S(t,k)$ (if any) to be broadcast at that time on that channel.  The
goal is to minimize the cost of the schedule, denoted $\COST(S)$ and defined as
the \emph{expected response time} plus the {\em broadcast cost} of $S$.

For a finite schedule $S$, the expected response time of $S$, denoted
$\ART(S)$, is defined as follows.  At each time unit, each message is requested
by some client with probability $p_i$.  Once a message is requested, the client
waits until the next time at which the message is scheduled on any channel (or
the end of the schedule, whichever comes first).  $\ART(S)$ is defined to be
the expected waiting time for a random request at a random time.  The broadcast
cost of $S$, denoted $\BC(S)$, is defined to be the total cost of scheduled
messages, divided by the length of the schedule.

Throughout the paper, if any real-valued function $f$ is defined with respect
to finite schedules, then we implicitly extend it to any infinite schedule $S$
as follows: $f(S) = \limsup_{n\rightarrow\infty} f(S_n)$, where $S_n$ denotes
$S$ restricted to the first $n$ time slots.  Thus, the above definitions of
expected response time and broadcast cost implicitly extend to infinite
schedules.  All of the infinite schedules considered in this paper will be
periodic, in which case this extension is particularly simple.

The data broadcast problem and special cases were studied in
\cite{A87,AW85,AW87,AGH98,BBNS98,G83,KS99,S00,VH97}.  Works studying 
applications and closely related problems include
\cite{A98,BC96,CC93,HW63,HR87,HMRTV89,IVB94,IR84,KL98,KZ98,S00,SL96,ST97,TX96,V94}.
Some of the above works study the generalization allowing messages to
have arbitrary lengths, which we do not consider here.

Ammar and Wong \cite{AW85,AW87} proved that there always exists an
optimal infinite schedule with finite period.  They also formulated a
natural relaxation of the problem that gives an explicit lower bound
on the optimum; the performance guarantee in this paper is proven with
respect to that lower bound.  More recently, constant-factor
polynomial-time approximation algorithms have been shown
\cite{AGH98,BBNS98}, the best to date being a $9/8$-approximation
\cite{BBNS98}.  Although the problem itself is not known to be
$\NP$-hard, several variants are known to be \cite{BBNS98,KS99,S00}.

Khanna and Zhou \cite[\S1.2]{KZ98} state that it is unknown whether the problem
is MAX-SNP hard, even when $W=1$ and without broadcast costs.  In this paper,
we show that it is not (unless P=NP).  We present the first deterministic
polynomial time approximation scheme for the problem, assuming the $W$ and each
cost is bounded by a constant.  By ``polynomial time'', we mean that the time
taken to output the finite description of the infinite schedule is polynomial
in the number of messages $m$ in the input.

\section{Summary of Approach}
Our algorithm is based on a simple new observation that works for a special
case of the problem.  We use fairly technical but to some extent standard
techniques to extend it to the general case.  We sketch the idea here,
glossing over a fair amount of technical detail.

Ammar and Wong \cite{AW85,AW87} relax the optimization problem by
allowing messages to (a) be scheduled at non-integer times and (b) to
{\em overlap}, while still insisting that the total {\em density} of
the scheduled messages is at most $W$, the number of channels (the
extension to the multiple channel case is due to~\cite{BBNS98}).  The
density of a message (or set of messages) is the total number of
scheduled times, divided by the length of the schedule.  Standard
calculus yields a solution to this relaxed problem.  The solution
specifies for each message $M_i$ a {\em density} $d^*_i$, meaning that
the message should be scheduled every $\tau^*_i = 1/d^*_i$ time units.

Ammar and Wong describe the following simple randomized rounding
algorithm for producing a real schedule: {\em For $t=1,2,\ldots$, for
$k=1\ldots W$, choose a single message $M_i$ randomly so that
$\Pr\{M_i \mbox{ selected}\}$ is $d^*_i/W=W/\tau^*_i$; schedule $M_i$
in schedule slot $S(t,k)$.}  They observe that the expected waiting
time for a random request for $M_i$ is essentially $\tau^*_i$ in this
schedule.  Since the expected waiting time in the relaxed schedule is
essentially $\tau^*_i/2$ (because an average request falls midway
between two successive broadcasts of $M_i$), this yields a
2-approximation w.r.t.\ expected response time. Since the expected
broadcast cost of $S$ is the same as the broadcast cost of the relaxed
solution, the algorithm is a $2$-approximation algorithm w.r.t.\ the
total cost.  Ammar and Wong also describe a greedy algorithm that
Bar-Noy, Bhatia, Naor and Schieber generalize in~\cite{BBNS98} to the
multiple channel case and prove to be essentially a derandomization of
the randomized algorithm, with the same performance guarantee.

\medskip\noindent {\bf Round-robin within groups.}  Since our goal is a PTAS,
we naturally group messages that are essentially equivalent (i.e.\ have
essentially the same cost and probability).  Our simple idea is the following
variation of Ammar and Wong's rounding scheme, which is most simply described
as follows: Schedule the messages as Ammar and Wong do, but then, {\em within
  each group, rearrange the messages so that they are scheduled in round-robin
  (cyclic) order.}  The broadcast cost is unchanged, but the expected response
time improves as follows.  Whereas before, a random request for a message $M$
in a group $G$ would have waited (in expectation) for $|G|$ messages from $G$
until finding its message, in the round-robin schedule, a random request for
$M$ will wait (by symmetry) for $(1+2+\cdots+|G|)/|G|) = (|G|+1)/2$ messages
from $G$.  That is, the expected wait in the round-robin schedule is
$(|G|+1)/(2|G|)$ times the expected wait in the Ammar-Wong schedule.  Since the
Ammar-Wong schedule has performance guarantee $2$, the round-robin schedule has
performance guarantee $\max_G (|G|+1)/|G| = 1 + 1/\min_G |G|$.  Thus, when the
groups are all large, the Ammar-Wong relaxation is essentially tight.

\medskip\noindent
{\bf Extending to the general case.}
Recall that for our purposes a group is a collection of messages
with approximately (w.r.t.\ $\epsilon$) the same probability and cost.
As long as each group has size at least $1/\epsilon$,
the round-robin schedule gives a $(1+\epsilon)$-approximation.

To extend to the general case, we show the following.
{\em Any} set of messages can be partitioned into three classes as follows:
\begin{enumerate}
\item[$A$] --- A constant number of {\em important} (high probability) messages.
\item[$B$] --- Messages belonging to {\em large groups}.
\item[$C$] --- Leftover messages, contributing {\em negligibly} to the cost.
\end{enumerate}
The basic intuition for the existence of this partition is that, due
to the rounding, the message-probabilities of the successive groups
decrease exponentially fast.  Thus, for all but a constant number of
groups (where the message-probability is high), either the group is
very large, or the total probability of the messages in the group is
very small.  Althouth the intuition is basic, obtaining the proof with
the appropriate parameters is is somewhat involved and delicate.

Once we have the partition, we proceed as follows:
\medskip
\\\noindent 1. Find the density $\alpha$ of messages in $A$ in a near-optimal
schedule of $A$ and $B$.
\\2. Compute an optimal ``short'' schedule $S_A$ of $A$ having density approximately $\alpha$.
\\3. Schedule the messages in $B$ in the slots not occupied by $A$,
using the group-round-robin algorithm.
\\4. ``Stretch'' the schedule, interspersing empty slots every $1/\epsilon$
time units, and schedule the messages for $C$ in these empty slots.
\medskip

Note that in order to ``cut and paste'' the schedules together, we have to
explicitly control the density of $A$ and $B$.  This in itself requires little
that is new.  The main new difficulty is the following.  In step 3, we are
using the round-robin algorithm to schedule $B$, but in a schedule that is
already partially filled by $A$.  For the analysis of the round-robin algorithm
to continue to approximately hold, we require that the empty slots in schedule
$S_A$ are sufficiently {\em evenly distributed} so that the scheduling of $B$
is not overly delayed at any time (cost increases quadratically with delay).

A-priori, imposing this additional requirement on $S_A$ might increase the cost
of $S_A$ too much.  To show that this is not the case, we show (using a
non-constructive probabilistic argument) that there is a schedule of $A$ that
has {\em constant-length period}, density approximately $\alpha$, and cost
approximately the cost of any optimal schedule of $A$ with density $\alpha$.
Since the period of this schedule is small, the empty slots are {\em necessarily}
evenly distributed.

The final output of the algorithm is a finite (size linear in the
input size) description from which an infinite schedule with
approximately optimal expected cost can be generated by a randomized
algorithm in an ``on-line'' fashion, where each step requires $O(W)$
time to schedule.

The running time of the various steps is as follows.  In step 1, only a
constant number of densities $\alpha$ need to be considered: we can try them
all and take the best.  For each $\alpha$, the time for the remaining steps is
as follows.  Step 2 can be done in constant time since the schedule we are
looking for has constant length.  Step 3 can be done in randomized time in the
size of the output.  Step 4 can also be done in randomized linear time in the
size of the output.

The final technical hurdle is showing that the algorithm can be
derandomized (extending the analysis of the greedy algorithm by
Bar-Noy, Bhatia, Naor and Schieber to this more complicated setting).
The resulting deterministic algorithm outputs a polynomial-length
schedule, the repetition of which gives the desired near-optimal
infinite schedule.

\section{Group round robin}
\label{sec:lb}
\label{sec:simplecase}

Let the set of messages $M$ be partitioned into groups
$G_1,\ldots,G_q$ where group $G_j$ has size $g_j$  every message in $G_j$
has the same probability $p_j$ and broadcast cost $c_j$.
Let $\alpha$ be the desired maximum density of $M$ in the schedule.
In this notation, Ammar and Wong's relaxation of the problem is:
$$
\LB(M,\alpha) = \left\{
        \begin{array}{c}
        \DS\min_{\tau>0} \DS\sum_{j=1}^q \frac{p_j g_j^2 \tau_j}2 
                                + \frac{c_j}{\tau_j}\\
        \text{\normalfont Subject to:} ~
                 \DS\sum_{j=1}^q \frac1{\tau_j} \leq \alpha W
        \end{array}\right.
$$

\begin{lemma}[Lower Bound \cite{AW85}]
The minimization problem~$\LB(B,\alpha)$ is a lower bound to the contribution of the
messages of~$B$ to the cost of any schedule~$S$ over $W$~channels, in
which $B$~has density~$\leq\alpha$.

The problem has a unique solution~$\tau^*$ satsifying: $g_j
\tau^*_j=\sqrt{(2c_j+\lambda^*)/p_j}$, for some $\lambda^*\geq0$. If
$\sum_{j\in B} g_j \sqrt{p_j/(2c_j)} \leq \alpha W$, then $\lambda^* =0$;
otherwise, $\lambda^*$~is the unique solution to: $\sum_{j\in B}
\sqrt{p_j/(2c_j+\lambda^*)} = \alpha W$.
\end{lemma}

\begin{lemma}[Randomized Approximation] \label{lem:rando}
In the setting of this section, the randomized
algorithm~\ref{algo:rando} constructs a one-channel schedule~$S$
whose cost satisfies:\\
\centerline{$
\expect[\COST(S)] = \DS{ \sum_{j=1}^q \left(
        p_j \frac{g_j(g_j+1)}2 \tau_j + \frac{c_j}{\tau_j}
        \right) - \frac12} $}\\
If $\tau$~is chosen in order to minimize~$\LB(M,1)$, then
algorithm~\ref{algo:rando} is a~$\max_j (1+1/g_j)$-approximation.
\end{lemma}

\begin{algorithm}[htb]
\caption{Group round-robin algorithm}\label{algo:rando}
\begin{algorithmic}
\IF{$\sum_{j=1}^q 1/\tau_j < 1$} 
        \STATE $\CD$ Add a dummy group~$G_0$
        with~$p_0=c_0=g_0=0$ and $1/\tau_0 = 1-\sum_{j=1}^q 1/\tau_j$.
\ENDIF
\FOR {$t=1..\infty$}
        \STATE $\CD$ Draw at random a group~$G_j$ with
        probability~$1/\tau_j$. Schedule in time slot~$t$, the next
        message of group $G_j$ in  Round Robin order, if $j\neq0$;
        and stay idle otherwise.
\ENDFOR
\end{algorithmic}
\end{algorithm}

\begin{proof}
A message of~$G_j$ is broadcast during a time slot with
probability~$1/\tau_j$, then the average density of the group~$G_j$
is then $1/\tau_j$. Then: $\expect[BC(S)]=\sum_j c_j/\tau_j$.\\
As explained above, a request for a message in~$G_j$ waits on average
$1/2$ until the end of the current time slot and then
$(g_j+1)/2$~broadcasts of a message in~$G_j$ on average. Then:
$\expect[\ART(S)]=\frac12+\sum_j p_j g_j \tau_j \frac{g_j+1}2$.\\
Finally if $\tau=\tau^*$, then since $\sum_j p_j g_j=1$ and
$\tau^*\geq1$, we get that: $\expect[\COST(S)] \leq
\sum_j(p_jg_j\frac{g_j+2}2\tau^*_j + c_j/\tau^*_j)$,
hence the claimed performance ratio.
\end{proof}

\begin{remark}
Note that the law of large numbers implies that the expected cost is
obtained with probability $1$.
\end{remark}


\section{Scheduling {\large $\mathbf A$} and {\large $\mathbf B$}}
\label{sec:criticalcase}

Next we treat the case where the set of the messages~$M$ can be partitioned
into two sets~$A$ and~$B$ such that
\begin{itemize}
\item $A$ consists of a constant number of messages
\item $B$ is partitioned into groups as in the previous section,
such that each group has size at least $\kappa(\epsilon) |A|^2$, 
where $\kappa=\kappa(\epsilon)$~will be
defined later.
\end{itemize}

Recall from the discussion in the introduction that the challenge
at this point is to show that there is a near-optimal schedule of $A$
with the appropriate density $\alpha$ and in which the empty slots
are relatively uniformly distributed.  If so, then we can find the desired
schedule for $A$ by exhaustive search, and then schedule $B$ into the empty
space in the schedule using the round-robin algorithm previously described.

To show the existence of the desired schedule for $A$, we show there is a
near-optimal schedule of $A$ with the appropriate density and with {\em
  constant period} (independent of $\alpha$).
\begin{lemma} \label{lem:boundedperiod}
Given a set of messages~$A$, with cost at most $\CC$, some
constant $0<\epsilon<1$ and a density~$0<\alpha<1$, there exists a
periodic schedule~$S$ satsifying:
\begin{enumerate}
\item   The density of empty slots is $S$ is approximately~$(1-\alpha)$:\\
$1-\alpha(S) \geq (1-\epsilon) (1-\alpha)$, ~and~
$\alpha(S)\in]0,1[$
\item   The cost of $S$ is approximately optimal:\\
\centerline{$\COST(S,A) \leq (1+\epsilon)\OPT(A,\alpha) + \epsilon/2$}
\item   The period $T_S$ of $S$ can be  bounded:\\
\centerline{$T_S\leq \cfrac{40 \ln (1+4/\epsilon)}{\epsilon^4(1-\epsilon/6)} \max(\CC,1)\cdot |A|^2$}
\end{enumerate}
\end{lemma}

\begin{proofSketch} 
Our proof uses the probabilistic method. The main new, simple idea, is
in the construction, which efficiently smoothes the cost function by
erasing its possible wide variations over time in the particular
schedule under study.

 Let~$T$ be a parameter to be determined later.  Let~$S^*$ be a periodic schedule of~$A$ with
 density~$\alpha$ and which is nearly optimal: basically,
 $\COST(S^*,A)\leq (1+\epsilon)\OPT(A,\alpha)$.  Let~$T^*$ denote the period
of~$S^*$, which w.l.o.g. is a multiple of $T$.

From~$S^*$, construct another periodic 
schedule~$S_2$ by inserting in~$S^*$, every $T$~steps from a random
starting point,
all the messages of~$A$ in a fixed order. $S_2$~is thus structured
into blocks of length~$T+|A|$. Let~$S_n$ be the random schedule
obtained by concatenating $n$~blocks chosen at random from~$S_2$. For
suitable values of~$T$ and~$n$, we can prove that with positive
probability, $S_n$~satisfies the first two statements of the
Lemma. The period of~$S_n$ is clearly~$n(T+|A|)$, which
together with the choice of $T$ gives the third
statement of the Lemma.
\end{proofSketch}

\begin{notation}
Let~$T(\epsilon) = \cfrac{40\ln (1+4/\epsilon)}
{\epsilon^4(1-\epsilon/6)} \max(\CC,1)$ denote the bound of the lemma for 
$\hbox{period}(S)/|A|^2$. Parameter $\kappa$ will be defined as~$\kappa(\epsilon)
\eqdef 2 W T(\epsilon)/\epsilon$.
\end{notation}


The algorithm for scheduling $A\cup B$ is given below, and is
an $\epsilon$-approximation for the cases studied in this section.
Since this is the critical case, the analysis is promoted
from a lemma to a proposition.
\begin{proposition} \label{pro:A+B}
Let $\epsilon < 1/7$. In the setting of this section,
 Algorithm~\ref{algo:A+B} yields a
schedule~$S$, such that: $$
\expect[\COST(S)] \leq (1+5\epsilon) \OPT(A\cup B)
$$
\end{proposition}

\begin{algorithm}[htb]
\caption{Scheduling~$A$ and~$B$} \label{algo:A+B}
\begin{algorithmic}
\FOR{$x=1..(W\cdot T(\epsilon)|A|^2-1)$} 
    \STATE $\CD$ Compute an optimal periodic schedule~$S_\alpha$
    of~$A$ with density~$\alpha= x/(W\cdot T(\epsilon) |A|^2)$ and
    period~$\leq T(\epsilon)|A|^2$.
\ENDFOR
\STATE  $\CD$ Choose $\alpha_0$ which minimizes:\\      
        \centerline{$\COST(S_{\alpha_0},A) + \LB(B,1-\alpha_0)$}
\STATE  $\CD$ Compute the $\tau^*$ that minimizes $\LB(B,1-\alpha_0)$;
        Add a dummy group~$G_0$ with $p_0=c_0=0$ and $\tau^*_0$, such
        that: $1/\tau^*_0 = (1-\alpha_0) W-\sum_{j\in B} 1/\tau^*_j$.
\ENSURE
\FOR{$t=1..\infty$}
        \STATE $\CD$ Schedule during time slot~$t$, the same
        messages of~$A$ on the same channels, as in~$S_{\alpha_0}$.
        \FORALL{empty slot~$s$ during time slot~$t$} 
                \STATE $\CD$ Draw a group~$G_j$ of~$B$ with
                probability~$1/(\tau^*_j(1-\alpha_0) W)$. Schedule in
                slot~$s$, the next message of~$G_j$ in Round Robin
                order, if~$j\neq0$; and stay idle otherwise.
        \ENDFOR
\ENDFOR
\end{algorithmic}
\end{algorithm}

\begin{proofSketch}
The proof works in two steps:
\begin{enumerate}
\item   \label{pf:stepANA1} Scheduling the messages of~$B$ with the 
randomized algorithm~\ref{algo:rando} in the empty slots achieves a
good approximation of $\COST(S_{\alpha_0},A)+\LB(B,1-\alpha_0)$ (Using
the mapping lemma~\ref{lem:mapping}).
\item   \label{pf:stepOPT1} $\COST(S_{\alpha_0},A)+\LB(B,1-\alpha_0)$ 
is a good approximation of the optimal cost (Using
Lemma~\ref{lem:boundedperiod}).
\end{enumerate}
\end{proofSketch}


\section{{\large $\mathbf C$} --- The negligible messages}
\label{sec:negligible}
In this section, to show how to incorporate the ``negligible'' messages into
the schedule.  We assume that the set of messages~$M$ is partitioned into two
sets~$AB$ and~$C$, where $C$ has a ``negligible contribution'' to the cost.
(This section can be skipped by the reader who is in a hurry).

\begin{definition} 
A subset of messages~$C\subseteq M$ has \emph{negligible contribution}
if its contribution to the lower bound is $O(\epsilon)$, when it is 
scheduled on one channel with density~$O(\epsilon/\CC)$, {\ie}:\\
\centerline{$\LB_{W=1}(C,\epsilon/(10\CC )) \leq 3\epsilon\OPT(M)/10$}
The constants~$1/10$ and~$3/10$ are arbitrary and are chosen in order
to improve readability in the following results.
\end{definition}

Basically, a subset of messages~$C$ is negligible if its
contribution to the cost is small, in the schedule
constructed by inserting its messages are
inserted from time to time (every $O(1/\epsilon )$ steps)
 into a schedule of the rest of the
messages.

\begin{lemma} \label{lem:scheduleC}
Consider a set of messages $M$, partitioned into two sets~$AB$
and~$C$, where $C$ has a negligible contribution to the cost. Then,
for any schedule~$S$ of~$AB$, one can construct in linear time a
random schedule~$S'$ of~$M$, such that: $$
\begin{array}{l}
\expect[\COST(S')] \\
~~~~~~~~~~~\leq (1+\frac{\epsilon}{10})\COST(S,AB)+ \frac{9\epsilon}{10}\OPT(M)
\end{array}
$$ 
If $\COST(S,AB)\leq\OPT(M)$, this is a $(1+\epsilon)$-approximation.
\end{lemma}

\begin{proof}
We first construct from $S$ another schedule $S_1$ of $AB$  by inserting 
an empty slot, on all the channels, every $10\CC/\epsilon-1$~slots,
starting at a random point in~$\{0,\ldots,10\CC/\epsilon-2\}$. The
stretching lemma~\ref{lem:stretching} ensures that:
$\expect[\COST(S_1)] \leq (1+\epsilon/10)\COST(S,AB)$. Let $\tau^*$ be
the solution to the minimization
problem~$\LB_{W=1}(C,\epsilon/(10\CC))$. We obtain~$S'$ by scheduling
the messages of~$C$ on  the {first} channel
in the empty slots of~$S_1$, according to the randomized 
algorithm~\ref{algo:rando}
with~$\tau=10\tau^*\CC/\epsilon$. Lemma~\ref{lem:rando} and the scaling
lemma~\ref{lem:scaling} ensure that the expected contribution of~$C$
is bounded by~$3\LB(C,\epsilon/(10\CC))\leq9\epsilon\OPT(M)/10$.
\end{proof}

\begin{remark}
The algorithm above can easily be derandomized by trying all the
starting point and choosing the one that minimizes the over cost for
the messages of~$AB$ and use the greedy algorithm~\ref{algo:greedy} to
schedule~$C$.
\end{remark}


\section{PTAS for Data Broadcast}
\label{sec:PTAS}
We now assume that we are in the general case.
The aim of the section is to prove the following theorem, which is the
main result of the paper. 
\begin{theorem}[PTAS] \label{thm:PTAS}
Given $\epsilon<1/7$ and a set~$M$ of messages, with message
costs bounded by $\CC$,  Algorithm~\ref{algo:PTAS} constructs in $O(m^2)$~time
a periodic schedule~$S$ with period~$\leq (m^2+m\max(1,\CC))/\epsilon$,
so that: $$
\COST(S)\leq (1+11\epsilon) \OPT(M)
$$
\end{theorem}
We will
first derive a PTRAS that will be derandomized in
Section~\ref{sec:derando}. 

\subsection{Randomized}
We now need to put together
the ideas developed for the special cases of the previous sections.
As a preliminary treatment, we use standard rounding techniques
to reduce the number of different messages.

%



\begin{lemma}[Rounding] \label{lem:rounding}
Without loss of generality, we can assume that the request
probabilities~$p_i$ are a multiple of powers of~$1/(1+\epsilon)$ and
the broadcast costs are multiples of~$\epsilon/W$:
\begin{eqnarray*}
p_i & = & r/(1+\epsilon)^j \text{, for some $j\geq 1$}\\ c_i & = &
k \cdot \epsilon/W \text{, for some $k\in\{0,\ldots, C\cdot
W/\epsilon\}$}
\end{eqnarray*}
where $1 < r \leq 1+\epsilon$.
\end{lemma}

\begin{proof}
Standard and omitted.
\end{proof}


The following lemma is the main tool for putting together the various
special cases studied so far, and is thus a key part of our construction.
We would like to claim that similar ideas could be applied to
other problems as well, however we were unable to abstract 
simple and general ideas from the technical proof. Perhaps, if one
believes that every approximation scheme rests on one ``structural lemma'',
 it can be seen as the structural lemma for this problem. 
\begin{lemma}[Partition] \label{lem:partition}
Given~$\epsilon>0$ and $\kappa>0$, one can construct, in linear time in
$m$, a partition of the groups~$(\Gjk)$, of messages with
probability~$r/(1+\epsilon)^j$ (where $r$ is the normalizing
constant such that $\sum_{\{j\geq1;k=0..\CC W/\epsilon\}}
r \gjk /(1+\epsilon)^j= 1$) and cost~$k\epsilon/W$, into three
sets~$A,B,C$ so that:
\begin{enumerate}
\item   The groups of $A$ have total size constant: 
$|A|\eqdef \sum_{\Gjk\in A}\gjk = O_{\epsilon,\kappa,\CC,W}(1)$, independent
of~$m$.
\item The groups of $B$ are all large:\\
\centerline{$\forall(\Gjk\in B), \gjk\geq \kappa |A|^2$}
\item The messages in $C$ have negligible contribution if they 
are scheduled rarely (with density~$O(\epsilon/\CC )$):\\
\centerline{$\LB_{W=1}(C,\frac{\epsilon}{10\CC}) 
\leq \frac{3\epsilon}{10}\OPT(M)$}
\end{enumerate}
\end{lemma}

\begin{proofSketch} 
Since the proof is rather technical, we will only in this extended
abstract give the construction of the partition into $A,B$ and $C$ in
the case when there are no costs ($\CC =0$) and there is only one
broadcast channel ($W=1$); this already contains the gist of the
proof.

Let $a=(1+\epsilon)^{-1}<1$.
In the case where there are no costs, the lower bound can be solved
explicitly (see \cite{AW85,BBNS98}) even when there is a density
constraint, to yield, for any subset $X$ of the message set:\\
\centerline{ $
\begin{array}{r@~c@~l}
LB(X,\alpha)&=
&\frac r{2\alpha } \bigl(\sum_{G_{j}\in X} g_{j} a^{j/2}\bigr)^2
\end{array}
$}

\paragraph*{\normalfont\itshape The construction} The construction
is best understood by referring to figure~\ref{fig:partition}.  We first 
deal with indices such that~$g_{j} \leq a^{-j/4}$. Let $j_0$ be some
constant to be defined later, and define $C_1 = \{ (j: j > j_0,~
g_{j} \leq a^{-j/4}\}$, and $A_1 = \{ j : j \leq j_0,~ g_{j}
\leq a^{-j/4}\}$. (One can observe already that since the contributions
of the messages of $C_1$ form the tail of a geometrically decreasing
series, they will be negligible, and so they will end up in $C$; moreover,
since $j$ and $g_j$ are both bounded for the definition of $A_1$, set
$A_1$ can only contain a small number of messages and so these messages
will end up in $A$).

We now consider the more delicate case of the
groups for which $g_{j} >a^{-j/4}$, for which we will need to use
the pigeon hole principle. We
partition their indices into $(20/\epsilon)$~blocks as
follows:
$$
\begin{array}{r@~c@~l}
\Lambda_1 & = & \{ (j,k) : \mu \leq j < \mu^2 \},
\smallskip \\
\Lambda_2 & = & \{ (j,k) : \mu^2 \leq j < \mu^3  \}, \ldots
\smallskip \\
\Lambda_{20/\epsilon} & = &  
\{ (j,k) : \mu^{20/\epsilon} \leq j
< \mu^{1+20/\epsilon} \},
\end{array}
$$
where $\mu$~is some constant to be defined later.  According
to~\cite{AW85}, we can then rewrite the lower bound on the expected
response time as $\sqrt{2 LB(M)/r} = \sum_j g_{j} a^{j/2}
\geq \sum_h \sum_{j\in\Lambda_h} g_{j} a^{j/2}$, and the pigeon
hole principle tells us that there exists at least one~$h$ such that
$\sum_{j\in \Lambda_h} g_j a^{j/2} \leq \frac{\epsilon}{20}
\sqrt{2 LB(M)/r}$. We now define $A_2 =
\{ j:j < \mu ^h,~ g_{j}>a^{-j/4} \}$, $C_2 = \Lambda_h$, and
$B = \{ j:j > \mu ^h,~ g_{j}>a^{-j/4} \}$.

Finally we set $A = A_1 \cup A_2$ and $C = C_1 \cup C_2$
 as shown on Figure~\ref{fig:partition}.

\begin{figure}[htb]
\hfill \input{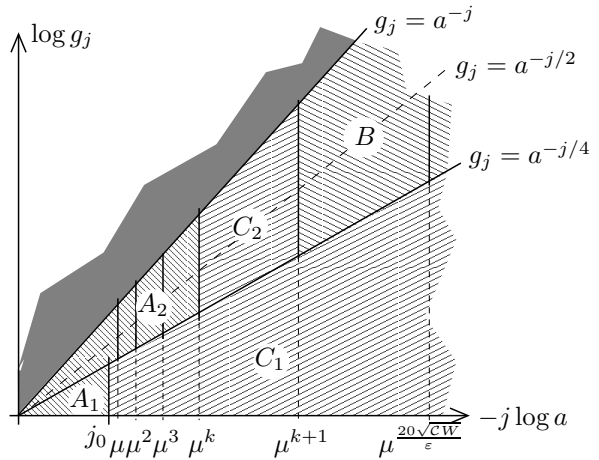} \hfill
\caption{The partition.} \label{fig:partition}
\end{figure}
\end{proofSketch}

It is now a simple matter to take our building blocks
and deduce a randomized approximation scheme for the general
Data Broadcast problem.

\begin{proposition}[PTRAS] \label{pro:PTRAS}
Given~$0<\epsilon<1/7$, the randomized algorithm~\ref{algo:PTRAS}
yields a random schedule~$S$ with cost: $$
\expect[\COST(S)] \leq (1+10\epsilon)\OPT(M)
$$
\end{proposition}

\begin{algorithm}[htb]
\caption{A PTRAS} \label{algo:PTRAS}
\begin{algorithmic}
\STATE $\CD$ Round the probabilities and costs of the messages in~$M$,
        and partition the set of messages~$M$ into three sets~$A,B,C$,
        according to Lemma~\ref{lem:partition} with
        $\kappa=\kappa(\epsilon)$.
\STATE $\CD$ Schedule $A$ and $B$ with algorithm~\ref{algo:A+B}.
\STATE $\CD$ Insert the messages of~$C$ into the schedule of~$A$ and~$B$, 
with the algorithm described in Lemma~\ref{lem:scheduleC}.
\end{algorithmic}
\end{algorithm}

\begin{proof}
Consider the rounded instance~$\dM$ of the set of messages. According
to the previous Proposition~\ref{pro:A+B} and
Lemma~\ref{lem:scheduleC}, we have:\\
\centerline{$
\expect[\COST(S)] \leq (1+\epsilon)(1+5\epsilon)\OPT(\dM)
$} But Lemma~\ref{lem:rounding} ensures that: \\
\centerline{$
\OPT(\dM)\leq(1+3\epsilon)\OPT(M)$}
which yields the result.
\end{proof}

\begin{note}
The insertion of~$C$ can be done at the same time than the broadcast
of~$A$ and~$B$ in Algorithm~\ref{algo:PTRAS}. 
\end{note}

\subsection{Derandomization}
\label{sec:derando}

\begin{figure}[htb]
\hfill \input{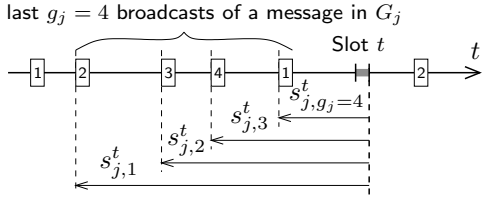} \hfill
\caption{Definition of the state at time slot $t$.}
\label{fig:greedystate}
\end{figure}

The PTRAS has one slight problem, namely, that it is not periodic,
hence may be somewhat awkward to implement in some settings. In this
section we derandomize it using greedy choices, and show how
to control the period of the resulting algorithm.
\begin{definition}[State] \label{not:greedystate}
We define the \emph{state} $(s^t_{j,k})_{1\leq j\leq q, 1\leq k\leq
g_j}$ at slot $t$ as the time period elapsed from the beginning of
the $k\th$ of the $g_j$ last broadcasts of group $G_j$ to the end of
slot $t$, as shown Figure~\ref{fig:greedystate}.
\end{definition}

\begin{lemma}[Derandomization of Algorithm~\ref{algo:rando}] \label{lem:greedy}
Given a set of messages partitioned into groups $\series G1q,$ of
size $g_j$, and a set of reals $\tau_j>0$ so that $\sum_j 1/\tau_j \leq
1$, the greedy algorithm~\ref{algo:greedy} yields a one-channel
schedule $S$ whose cost satisfies:$$
\COST(S) \leq \sum_{j=1}^q \left(
        p_j \frac{g_j(g_j+1)}2 \tau_j + \frac{c_j}{\tau_j}
        \right) -\frac12
$$
If $\tau$~minimizes~$\LB(M,1)$, we get a $\max_j(1+1/g_j)$-approximation.
\end{lemma}

\begin{algorithm}[htb]
\caption{Greedy Algorithm}\label{algo:greedy}
\begin{algorithmic}
\STATE $\CD$ Add a dummy group $G_0$, if needed.
\FOR{$t=1..\infty$}
        \STATE $\CD$ Let $(s_{j,k})$ be the state at time slot $t-1$.
        \STATE $\CD$ Let $j\in\{0,\ldots,q\}$ which minimizes:\\
        \centerline{$(c_j-p_j\tau_j\sum_{k=1}^{g_j} s_{j,k})$}
        \STATE $\CD$ Schedule during slot $t$, the next message of $G_j$ in
        the Round Robin order, if $j\neq0$, and stay idle otherwise.
\ENDFOR
\end{algorithmic}
\end{algorithm}

\begin{proofSketch}
The greedy choice at time slot~$t$ is made in order to minimize the
expected cost of the already allocated slots~$1,\ldots,(t-1)$, if the
schedule continues with the randomized algorithm~\ref{algo:rando}
after time~$t$; this property ensures that the greedy schedule is at
least as good as the randomized one.
\end{proofSketch}

The above greedy algorithm could conceivably have very large period.
The lemma below shows that we can truncate it so as to obtain
a periodic schedule of polynomial length.
\begin{corollary}[Greedy periodic schedule] \label{lem:periodicgreedy}
Given a set of messages partitioned into groups $\series G1q,$ of
size $g_j$,  a set of reals $\tau_j>0$ such that $\sum_j 1/\tau_j
\leq W$, and any~$T \geq (8m^2+(4\CC-1)m)$, 
Algorithm~\ref{algo:periodicgreedy} yields a one-channel schedule~$S$
with period~$(T+2m)$, whose cost is bounded by: $$
\COST(S) \leq \sum_{j=1}^q \left(
        p_j \frac{g_j(g_j+1)}2 \tau_j + \frac{c_j}{\tau_j}
        \right)
$$
\end{corollary}

\begin{algorithm}[htb]
\caption{A periodic greedy algorithm} \label{algo:periodicgreedy}
\begin{algorithmic}
\STATE $\CD$ Schedule during slot $t=1..m$ message $M_t$.
\STATE $\CD$ Execute the greedy algorithm during slots $t=(m+1)..(T+m)$.
\STATE $\CD$ Sort in increasing order the set $\{k\tau_j: 1\leq j\leq q; 
1\leq k\leq g_j\}$ and Schedule in slots $t=(T+m+1)..(T+2m)$ in order
of increasing $k\tau_j$, the $k\th$ message of group $G_j$ in the
Round Robin order.
\end{algorithmic}
\end{algorithm}

\begin{proof}
Omitted.
\end{proof}

Our main algorithm can now be found in Algorithm~\ref{algo:PTAS}.
\begin{algorithm}[htb]
\caption{The PTAS} \label{algo:PTAS}
\begin{algorithmic}[1]
\STATE  Round the probabilities and costs, and 
partition~$M$ into~$A,B,C$ as in the PTRAS.
\STATE Compute $\tau^*$ and the density $\alpha_0$ and periodic
schedule $S_{\alpha_0}$ of $A$ to minimizes $\LB(B,(1-\alpha_0)$, as in Algorithm 2.
\STATE Compute the greedy periodic schedule~$S_B$ of~$B$ 
with $\tau=(\tau^*(1-\alpha_0)W)$ and with
period~$\bigl\{T(\epsilon)|A|^2(1-\alpha_0)W(8m^2+(4\CC +1)m)\bigr\} =
\Theta(m^2)$.
\STATE Concatenate $(8m^2+(4\CC+1)m)$ periods of~$S_{\alpha_0}$ and 
map~$S_B$ into the empty slots in the natural order.
\STATE Compute the greedy periodic schedule~$S_C$ of~$C$ 
with $\tau=(10\CC\tau^*/\epsilon)$ where $\tau^*$~minimizes
$\LB(C,\epsilon/10\CC)$, and with period~$\bigl\{T(\epsilon)|A|^2(8m^2+(4\CC
+1)m)\epsilon/10\CC\bigr\} = \Theta(m^2)$.
\STATE Choose the best starting point in~$\{1,\ldots,\frac{10\CC}\epsilon-1\}$
and stretch the schedule of $A$ and $B$ by inserting a slot of~$S_C$
on the first channel every $(\frac{10\CC}\epsilon-1)$ and an empty slot
on the other channels at that time. Let $S$ be the
resulting schedule.
\STATE Choose the best starting point 
in~$\{1,\ldots,\frac{m^2+m\CC}\epsilon-m\}$ and construct~$S'$ by
stretching~$S$ by inserting the $m$~messages in fixed order on the
first channel every $(\frac{m^2+m\CC}\epsilon-m)$.
\STATE $S'$ is then structured into independent blocks of
length $\frac{m^2+m\CC}\epsilon$. The cheapest block~$S^*$ will be the
period of our approximation.
\end{algorithmic}
\end{algorithm}

\begin{proofOf}{Theorem~\ref{thm:PTAS}}
Theorem~\ref{thm:PTAS} is proved by analyzing the
algorithm~\ref{algo:PTAS}. The analysis is derived from the analysis
of the PTRAS. The six first steps are exactly the same, except that
the periodic greedy algorithm~\ref{algo:periodicgreedy} is used
instead of the randomized algorithm~\ref{algo:rando}. Since the
performance ratio in Algorithm~\ref{algo:periodicgreedy} is better,
the schedule~$S$ obtained Step~6 is at least as good, and is periodic
with period~$O(m^2)$: \\
\centerline{$
\COST(S) \leq (1+10\epsilon)\OPT(M)
$}

We finally reduce the period in Steps~7-8 by using  stretching
lemma~\ref{lem:stretching}, which ensures that at an increase of
$(1+O(\epsilon))$ of the cost, we can extract from~$S$ a block~$S^*$
with length $\leq\frac{m^2+m\max(1,\CC)}\epsilon$ and:\\
\centerline{$
\COST(S^*) \leq (1+11\epsilon)\OPT(M)
$}
\end{proofOf}


\section{Technical lemmas}\label{sec:technical}

The lemmas in this sections are useful for analyzing several of
our constructions. The stretching lemma states that changing a schedule
by inserting a few empty slots once in a while does not affect
the expected response time.

\begin{lemma}[Stretching] \label{lem:stretching}
Given a schedule~$S$ on $W$~channels of~$M$ and a positive
integer~$y$, let $\kappa \geq \frac{y^2+y}\epsilon-y$. Consider the
schedule~$S'$ obtained from~$S$ by inserting~$y$ empty slots just
before the time slots~$x,x+\kappa,\ldots,x+i\cdot\kappa,\ldots$, where
$x$~is a random time in~$\{1,\ldots,\kappa\}$. Then: $$
\expect[\ART(S')] \leq (1+\epsilon)\ART(S)
$$
\end{lemma}

\begin{proof}
Omitted.
\end{proof}

The scaling lemma is immediate.
\begin{lemma}[Scaling] \label{lem:scaling}
Given a set of messages $M$ and a schedule $S$, let $S_\alpha$ the
schedule obtained by scaling $S$ by a factor $1/\alpha$: $S_\alpha$
schedule at time $t/\alpha$ on some channel the same message as $S$ at
time $t$, and stays idle otherwise. Then:
\begin{eqnarray*}
\ART(S_\alpha,A) & = & \cfrac1\alpha\cdot\ART(S,A) \\
\BC(S_\alpha,A) & = &  \alpha \cdot \BC(S,A)
\end{eqnarray*}
\end{lemma}

\begin{proof}
Immediate.
\end{proof}


The mapping lemma is used for analyzing the effect of inserting
the messages from $B$ into the slots left empty in the 
density-constrained schedule of $A$; these slots may be spaced
irregularly.
\begin{lemma}[Mapping into reserved empty slots] \label{lem:mapping}
Given a set of messages~$M$, partitioned into groups of 
identical messages, such that all groups are larger than $TW$,
consider a one-channel schedule $S$ of $M$ scheduling each group in
Round Robin order, and a periodic sequence of reserved time-slots over
$W$~channels with density~$\alpha$ and period~$T$. Let~$S'$ be
the schedule obtained by mapping the schedule~$S$ into the reserved
empty slots from left to the right, then:
\begin{eqnarray*}
\ART(S',A) & \leq & \cfrac1{\alpha W} \cdot \ART(S,A) 
                        + T \sum_{M_i\in A} p_i \\
\BC(S',A) & = &  \alpha W \cdot\BC(S,A)
\end{eqnarray*}
\end{lemma}

\begin{proof}
Omitted.
\end{proof}

\begin{corollary}[Case of large groups]
In the case where $M$~is partitioned into  groups of size~$\geq
2T/\epsilon$, $S'$~has a cost bounded by: \smallskip \\
$~~~\COST(S',A) \leq$\\
$~~~~~~~~~~~~~~(1+\epsilon) \biggl\{ \cfrac1{\alpha W}\cdot 
\ART(S,A) + \alpha W \cdot \BC(S,A) \biggr\} $
\end{corollary}

\begin{proof}
Simple calculation.
\end{proof}

\end{document}